\documentclass[%
 reprint,
superscriptaddress,
 amsmath,amssymb,
 aps,
pra,
]{revtex4-2}
\usepackage[dvipsnames]{xcolor}
\usepackage[english]{babel}
\usepackage{textcomp}
\usepackage{amsmath}
\usepackage{array, multirow, bigdelim, makecell, booktabs} 
 
\makeatletter
\def\bbl@set@language#1{%
  \edef\languagename{%
    \ifnum\escapechar=\expandafter`\string#1\@empty
    \else\string#1\@empty\fi}%
  \@ifundefined{babel@language@alias@\languagename}{}{%
    \edef\languagename{\@nameuse{babel@language@alias@\languagename}}%
  }%
  \select@language{\languagename}%
  \expandafter\ifx\csname date\languagename\endcsname\relax\else
    \if@filesw
      \protected@write\@auxout{}{\string\select@language{\languagename}}%
      \bbl@for\bbl@tempa\BabelContentsFiles{%
        \addtocontents{\bbl@tempa}{\xstring\select@language{\languagename}}}%
      \bbl@usehooks{write}{}%
    \fi
  \fi}
\newcommand{\DeclareLanguageAlias}[2]{%
  \global\@namedef{babel@language@alias@#1}{#2}%
}
\makeatother

\DeclareLanguageAlias{en}{english}

\usepackage{graphicx}
\usepackage{amsmath}
\usepackage{array}
\usepackage{amssymb}
\usepackage{dcolumn}
\usepackage{bm}
\let\savecorresponds\corresponds
\let\corresponds\relax
\usepackage{mathabx}
\usepackage{enumitem} 
\usepackage{tabularx}
\usepackage{soul}
\usepackage{ulem}
\let\corresponds\savecorresponds
\usepackage{hyperref}
\usepackage[utf8]{inputenc}

\def\pd2v#1#2#3{\frac{\partial^2 #1}{\partial #2 \partial #3}}

\def \2x2mat#1#2#3#4{
\left( \begin{array}{cc}
#1 &  #2 \\  #3 &  #4
\end{array} \right)
}

\begin{document}

\preprint{APS/123-QED}

\title{Quantifying high-dimensional spatial entanglement with a single-photon-sensitive time-stamping camera}

\author{Baptiste Courme}
\affiliation{Sorbonne Université, CNRS, Institut des NanoSciences de Paris, INSP, F-75005 Paris, France\
}
\affiliation{Laboratoire Kastler Brossel, ENS-Universite PSL, CNRS, Sorbonne Universite, College de France, 24 rue Lhomond, 75005 Paris, France\
}%
\author{Chloé Vernière}
\affiliation{Sorbonne Université, CNRS, Institut des NanoSciences de Paris, INSP, F-75005 Paris, France\
}
\author{Peter Svihra}
\affiliation{Czech Technical University, 115 19 Prague 1, Czech Republic\
}%
\affiliation{CERN, 1211 Geneva, Switzerland\%
}%
\author{Sylvain Gigan}
\affiliation{Laboratoire Kastler Brossel, ENS-Universite PSL, CNRS, Sorbonne Universite, College de France, 24 rue Lhomond, 75005 Paris, France\
}%
\author{Andrei Nomerotski}
\affiliation{Brookhaven National Laboratory, Upton, NY, 11973, USA\
}%
\author{Hugo Defienne} \email[Corresponding author: ]{hugo.defienne@insp.upmc.fr}
\affiliation{Sorbonne Université, CNRS, Institut des NanoSciences de Paris, INSP, F-75005 Paris, France\
}

\date{\today}
\begin{abstract}
High-dimensional entanglement is a promising resource for quantum technologies. Being
able to certify it for any quantum state is essential. However, to date, experimental entanglement certification methods are imperfect and leave some loopholes open. Using a single-photon sensitive time-stamping camera, we quantify high-dimensional spatial entanglement by collecting all output modes and without background subtraction, two critical steps on the route towards assumptions-free entanglement certification. We show position-momentum Einstein-Podolsky-Rosen (EPR) correlations and quantify the entanglement of formation of our source to be larger than $2.8$ along both transverse spatial axes, indicating a dimension higher than $14$. Our work overcomes important challenges in photonic entanglement quantification and paves the way towards the development of practical quantum information processing protocols based on high-dimensional entanglement.
\end{abstract}

\maketitle

High-dimensional entangled states offer several advantages over qubits. They have a greater information capacity~\cite{bechmann-pasquinucci_quantum_2000}, an improved computational power~\cite{reimer_high-dimensional_2019} and increased resistance to noise and losses~\cite{ecker_overcoming_2019}. They are also good candidates for use in device-independent quantum communication protocols~\cite{acin_device-independent_2007}. In particular, pairs of photons entangled in their transverse spatial degree of freedom have such a high-dimensional structure. When harnessed in the continuous bases of position and momentum~\cite{walborn_spatial_2010}, these states can lead to many applications, such as quantum key distribution~\cite{almeida_experimental_2005}, continuous-variable quantum computation~\cite{tasca_continuous-variable_2011} and quantum imaging~\cite{moreau_imaging_2019}.

To fully take advantage of high-dimensional entanglement, it is essential to certify its presence using as few measurements as possible and without making any assumptions. The most common experimental method consists in analysing the quantum state sequentially using single-outcome projective measurement~\cite{mair_entanglement_2001}. In particular, this method has been used to measure high-dimensional entanglement using different approaches, such as measuring Einstein-Podolsky-Rosen (EPR)-type correlations~\cite{howell_realization_2004,achatz_certifying_2022}, using compressive sensing~\cite{schneeloch_quantifying_2019}, quantum steering~\cite{designolle_genuine_2021}, or by detecting photons in mutually unbiased bases (MUB)~\cite{bavaresco_measurements_2018} . 

However, this approach has major shortcomings. First, it is time-consuming. For example, in the case of a bipartite state with local dimension $d$, it requires performing at least $2 d^2$ measurements (e.g. using two MUBs), making this task impractical in high dimensions and effectively limiting the key rate in quantum communications scenarios~\cite{valencia_high-dimensional_2020}. Second, it necessarily leaves the fair-sampling loophole open. Indeed, for the latter to be closed, one must ensure that all possible output states are measured simultaneously~\cite{christensen_detection-loophole-free_2013}.

Due to the limitations of single-outcome projective measurements, researchers are exploring new approaches to detect all outputs simultaneously rather than measuring them one at a time. One promising path is to use single-photon-sensitive cameras to detect photons in all spatial modes in parallel. High-dimensional spatial entanglement was thus recently measured using Electron Multiplied Charge Coupled Device (EMCCD) cameras~\cite{tasca_imaging_2012,moreau_realization_2012,moreau_einstein-podolsky-rosen_2014,reichert_massively_2018}, intensified scientific Complex Metal-Oxyde (i-sCMOS) cameras~\cite{dabrowski_einsteinpodolskyrosen_2017,dabrowski_certification_2018} and Single-Photon Avalanche Diode-CMOS sensor (SPAD camera)~\cite{ndagano_imaging_2020,eckmann_characterization_2020}. However, all of these methods require to post-process the recorded data, which necessarily opens a loophole in the certification protocol~\cite{zhu_is_2021}. Specifically, none of the results reported in Refs.~\cite{tasca_imaging_2012,moreau_realization_2012,moreau_einstein-podolsky-rosen_2014,reichert_massively_2018,dabrowski_einsteinpodolskyrosen_2017,dabrowski_certification_2018,ndagano_imaging_2020,eckmann_characterization_2020} were sufficient to violate the criterion of entanglement without modeling and subtracting accidental coincidences. 
 
The presence of such a high rate of accidental coincidences is due to the technical imperfections of these camera technologies, such as their low temporal resolution (EMCCD), poor quantum efficiency (SPAD array) or the presence of noise (CMOS)~\cite{moreau_imaging_2019}. In our work, we quantify high-dimensional spatial entanglement without post-processing the data using a recently developed single-photon-sensitive time-stamping camera~\cite{nomerotski_imaging_2019}.
We achieve violation of an EPR-type criterion and measure an entanglement of formation (Eof) of $2.81(3)$ and $3.02(3)$ along the $X$ and $Y$ transverse spatial axes, respectively, thus showing an entanglement dimension higher than $14$.
\begin{figure}[ht]
\centering
\includegraphics[width=1\columnwidth]{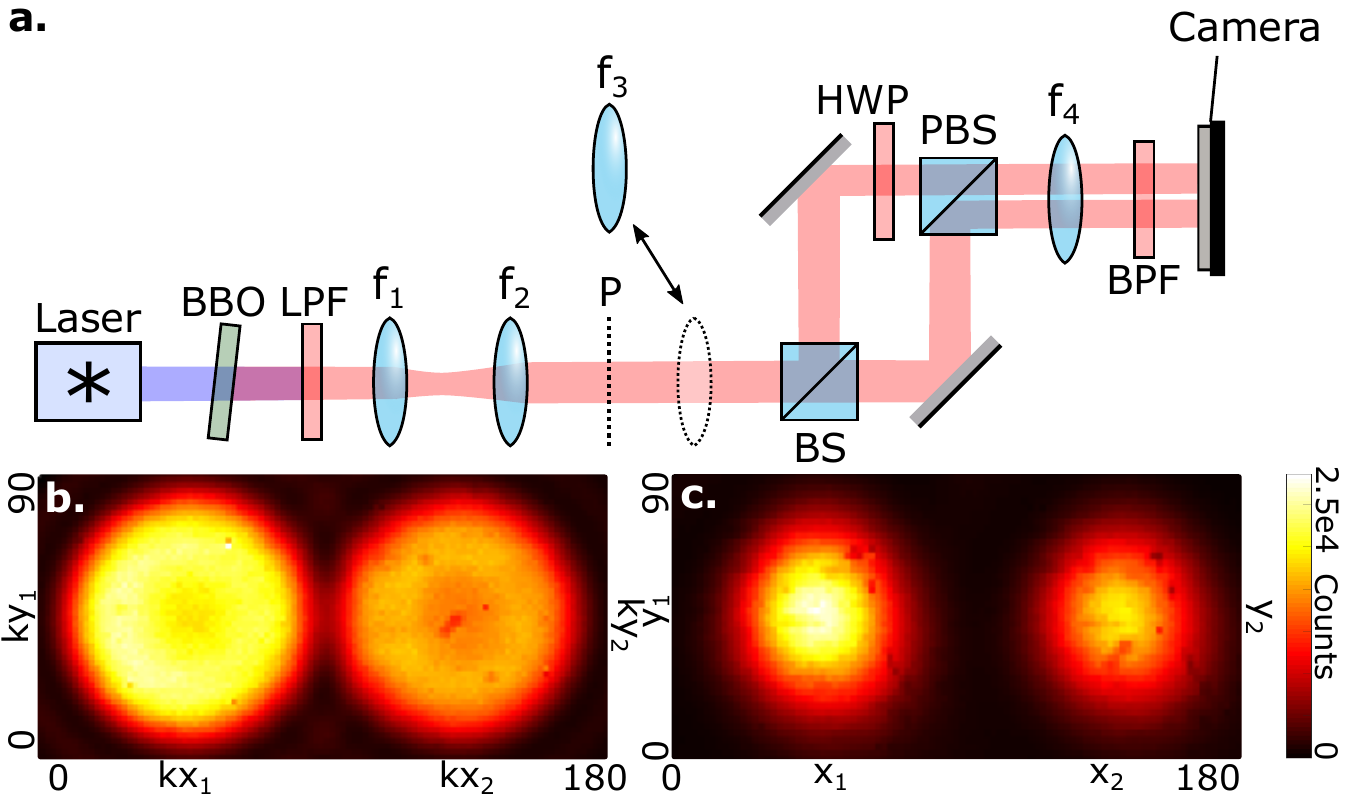}
\caption{\label{Figure1}\textbf{Experimental setup.} \textbf{(a)} A vertically-polarized continuous-wavelength collimated $405$nm laser diode illuminates a $0.5$mm-long $\beta$-barium-borate (BBO) crystal to produce spatially entangled photon pairs by type-I spontaneous parametric down conversion (SPDC). Crystal surface is imaged and magnified ($\times 2$) into an intermediate optical plane $P$ (dashed line) by a set of lenses $f_1$-$f_2$. A beam splitter (BS) separates photon-pairs beam into two that are re-positioned side-by-side using a half-wave plate at $45^\circ$ and a polarizing beam splitter (PBS). In the far-field (FF) configuration, the crystal is Fourier imaged onto the camera by lens $f_4$. In the near-field (NF) configuration, the crystal surface is imaged onto the camera using lenses $f_3=30$mm and $f_4=150$mm. The camera is composed of an image intensifier and a time-stamping camera $Tpx3Cam$. Long pass (LPF) and band pass (BPF) filters block the pump after the crystal to select near-degenerate photon pairs at $810 \pm 5$ nm. \textbf{(b)} and \textbf{(c)} show intensity images measured on the camera in the FF and NF configurations, respectively. For clarity, the spatial positions with an index '1' correspond to pixels on the left side of the camera and those with an index '2' to pixels on the right side. Spatial axes units are in pixels and intensity is in photon counts. Acquisition time is $200$ seconds.}
\end{figure}

Figure~\ref{Figure1}.a shows the experimental setup. Spatially entangled photon pairs are produced via type-I spontaneous parametric down conversion (SPDC) in a $\beta$-barium-borate (BBO) crystal using a $405$-nm continuous-wave pump laser. After magnification, a beam splitter divides the optical path into two paths imaged onto two separate parts of a  single-photon-sensitive time-stamping camera. In our work, two distinct imaging configurations are used. To detect photons in the momentum basis, the Fourier plane of the crystal surface is imaged onto the camera by the lens $f_4$ (far-field (FF) configuration). To detect photons in the position basis, we insert the lens $f_3$ to form a 4f-telescope (i.e. $f_3$ and $f_4$) and image the output surface of the crystal onto the camera (near-field (NF) configuration). Figures~\ref{Figure1}.b and c show intensity images measured by the camera using the FF and NF configurations, respectively.  

A central element in our experiment is the single-photon-sensitive time-stamping camera. It consists of a \textit{Tpx3Cam} camera (Amsterdam Scientific Instruments) and an image intensifier (Photonis). In contrast to the CCD and CMOS cameras, which acquire the images frame by frame, the \textit{Tpx3Cam} is an event-driven camera, which time-stamps the single photons continuously. After performing an acquisition and after post-processing of the raw data~\cite{ianzano_fast_2020}, a list of all the recorded single photons together with their time-stamps and spatial coordinates is formed. The list is then filtered to keep only coincidence events i.e. pairs of photons detected within a $6$ns time-window. Similar approaches were used in recent quantum and correlation imaging experiments~\cite{ianzano_fast_2020,svihra_multivariate_2020,zhang_high_2021,gao_manipulating_2022,zhang_ray_2022,vidyapin_characterisation_2023}. In our work, we use it to retrieve the full spatial two-photon joint probability distribution (JPD) of detected photons. In the FF configuration, it is noted $\Gamma(k_{x_1},k_{y_1},k_{x_2},k_{y_2})$ and represents the joint probability of detecting a photon at spatial position $(k_{x_1},k_{y_1})$ (left half of the camera) and a photon at position $(k_{x_2},k_{y_2})$ (right half). In the NF configuration, $\Gamma({x_1},{y_1},{x_2},{y_2})$ is  the joint probability of detecting a photon at position $({x_1},{y_1})$ (left half) and a photon at position $({x_2},{y_2})$ (right half).
\begin{figure*}[ht]
\centering
\includegraphics[width=1\textwidth]{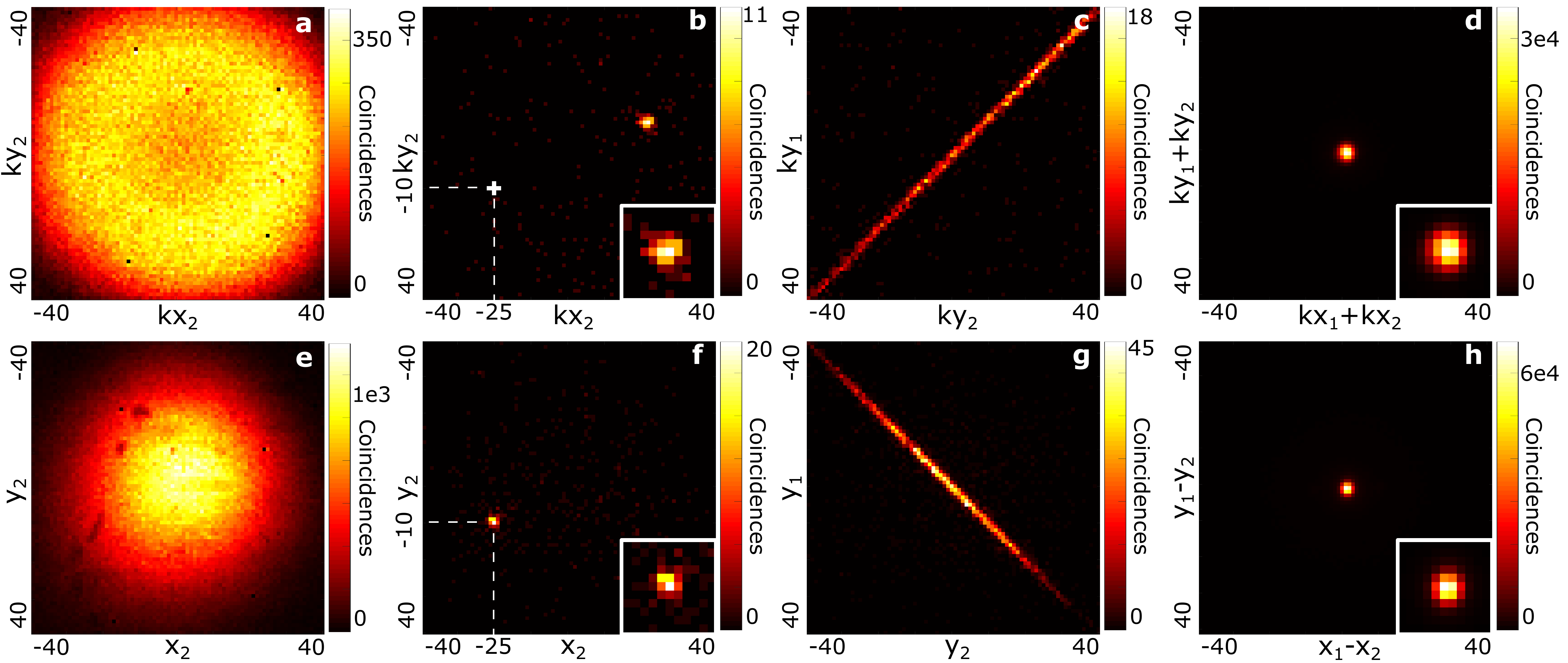}
\caption{\label{Figure2}\textbf{Bi-dimensional projections of the measured spatial joint probability distribution (JPD)} \textbf{(a)} and \textbf{(e)} Marginal probability distributions $\Gamma_m(k_{x_2},k_{y_2})$ and $\Gamma_m(x_2,y_2)$ of detecting one photon of the pair in the FF and NF configurations, respectively. \textbf{(b)} and \textbf{(f)} Conditional probability distributions $\Gamma(k_{x_2},k_{y_2}|k_{x_1},k_{y_1})$ and $\Gamma({x_2},{y_2}|{x_1},{y_1})$ of detecting one photon of a pair when its twin was detected at pixel $(k_{x_1},k_{y_1})=({x_1},{y_1})=(-25,-10)$ (white dashed lines) in the FF and NF configurations, respectively.  \textbf{(c)} and \textbf{(g)} Joint probability distribution $\Gamma(k_{y_2},k_{y_1},k_{x_2},k_{x_1})$ and $\Gamma({y_2},{y_1},{x_2},{x_1})$ of detecting photon pairs on the central column of the sensor i.e. $(k_{x_1},k_{x_2}) = (0,0)$ and $(x_1,x_2) = (0,0)$, in the FF and NF configurations, respectively. \textbf{(d)} Sum-coordinate projection $\Gamma_+(k_{x_1}+k_{x_2},k_{y_1}+k_{y_2})$ of the JPD measured in the FF configuration. \textbf{(h)} Minus-coordinate projection $\Gamma_-({x_1}-{x_2},{y_1}-{y_2})$ of the JPD measured in the NF configuration. Definitions of $\Gamma_m$, $\Gamma_-$ and $\Gamma_+$ are provided in the supplementary document. Inserted images are zooms around the peaks ($10\times 10$ pixels). Acquisition time is $200$ seconds in each configuration. Spatial axes units are in pixels. The measured JPD are not normalized and their units are numbers of coincidences. Totals of $1.4.10^6$ and $2.1.10^6$ coincidences were detected by the sensor during acquisition in the FF and NF configuration, respectively.}
\end{figure*}
 Figures~\ref{Figure2}.a-h show bi-dimensional projections of measured JPDs. In the momentum basis (FF configuration), JPD projections in Figures~\ref{Figure2}.b.c.d show that photons are anti-correlated on the camera i.e. when a photon is detected at position $(k_{x_2},k_{y_2})$, its twin is detected with a high probability at $(k_{x_1},k_{y_1}) = (-k_{x_2},-k_{y_2})$. In the position basis (NF), we observe that photons are strongly correlated i.e. they are always detected next to each other (Figs.~\ref{Figure2}.f.g.h). As position and momentum bases are mutually unbiased, the existence of strong spatial correlations suggest the presence of spatial entanglement~\cite{spengler_entanglement_2012}. Formally, its presence can be demonstrated using separability criteria. In our work, we use the EPR-Reid criterion~\cite{reid_colloquium_2009} based on the  inequality
\begin{equation}
\label{epr_criterion}
\Delta_{min}[x]\Delta_{min}[k_x] \geq \frac{1}{2}, 
\end{equation}  
where $\Delta_{min}[x]$ and $\Delta_{min}[k_x]$ are the minimum inferred uncertainties for position and momentum measurements, respectively. The same inequality exists also for the $Y$-axis after substituting $x \rightarrow y$ and $k_x \rightarrow k_y$. Uncertainties are expressed from measurable quantities by the following definition:
\begin{equation}
\label{definition}
\Delta^{2}_{min}[x] = \int d x_1 d x_2 \Gamma_m(x_2) \Delta^2[x_1|x_2], 
\end{equation}
where $\Delta[x_1|x_2]$ is the uncertainty in detecting a photon at $x_1$ conditioned on a detection at $x_2$ and $\Gamma_m(x_2)$ is the marginal probability of detecting a photon at $x_2$. The same definition applies for $\Delta_{min}[y]$, $\Delta_{min}[k_x]$ and $\Delta_{min}[k_y]$. Conditional uncertainties are estimated from conditional probability distribution, such as those shown Figures~\ref{Figure2}.b.f, by Gaussian fitting~\cite{schneeloch_introduction_2016}. All uncertainty values are reported in Table~\ref{table_certification}. In particular, we find $\Delta_{min}[x] \Delta_{min}[k_x] = 0.0333(6)$ and $\Delta_{min}[y] \Delta_{min}[k_y] =  0.0366(6) $, showing clear violation of the inequality~\eqref{epr_criterion} along both spatial axes. This  shows the presence of EPR correlations, and thus spatial entanglement, in the detected quantum state. 
\begin{table}[ht]
    \begin{tabular}{ c c c }
        \hline
       Quantity  &  Values &    Product    \\
       \hline
         $\Delta[k_{x_2}|k_{x_1}]$ & $3.5(5).10^3$ m$^{-1}$ &  \\
         $\Delta[k_{y_2}|k_{y_1}]$ & $3.1(5).10^3$ m$^{-1}$ &  \\
        $\Delta[x_2|x_1]$ & $1.0(5).10^{-5}$ m &  \\
       $\Delta[y_2|y_1]$ & $9.7(9).10^{-6}$ m &  \\
       $\Delta_{min}[k_x]$ & $3.217(5).10^{3}$ m$^{-1}$ & \hspace{-0.5em}\rdelim\}{2}{*}[ $ 0.0333(6) $]  \\
       $\Delta_{min}[x] $ & $1.03(1).10^-5$ m &  \\
       $\Delta_{min}[k_y]$ & $3.351(4).10^{3}$ m$^{-1}$ &  \hspace{-0.5em}\rdelim\}{2}{*}[ $ 0.0366(6) $] \\
       $\Delta_{min}[y]$ & $1.09(1).10^{-5}$ m & \\
       $\Delta[k_{x_2}+k_{x_1}]$ & $3.82(4).10^3$ m$^{-1}$ & \hspace{-0.5em}\rdelim\}{2}{*}[ $ 0.0524(7) $]  \\
        $\Delta[y_2-y_1]$ & $1.28(2).10^{-5}$ m &  \\
              $\Delta[k_{y_2}+k_{y_1}]$ & $4.07(4).10^3$ m$^{-1}$ & \hspace{-0.5em}\rdelim\}{2}{*}[ $ 0.0450(7) $]  \\
       $\Delta[x_2-x_1]$ & $1.17(1).10^{-5}$ m &   \\
      Lower bound $E_x$ & $3.03(2)$ &  \\
      Lower bound $E_y$ & $2.81(2)$ &  \\
       Lower bound $d_x$ & $8.1(1)$ &  \\
       Lower bound $d_y$ & $7.01(9)$ &  \\
       \hline
    \end{tabular}
    \caption{Measured values of uncertainties, lower bounds of entanglement of formation and low bound of dimension. $\Delta[k_{x_2}|k_{x_1}]$, $\Delta[_{y_2}|k_{y_1}]$ are conditional uncertainties measured along both axes for $(k{x_1},k_{y_1})=(-25,10)$  i.e. widths of the peak shwon in Figure~\ref{Figure2}.b. $\Delta[x_2|x_1]$, $\Delta[y_2|y_1]$ are conditional uncertainties measured along both axes for $(x_1,y_1)=(-25,10)$  i.e. widths of the peak shown in Figure~\ref{Figure2}.d. $\Delta_{min}$ values were calculated using equations~\eqref{definition}. Their product value show violation of the EPR-Reid criterion. Lower bounds of the Eof and dimension of entanglement are obtained using~\eqref{eq_eof_bound}. Errors in the measured and calculated values are written between parenthesis and apply to the last digit of the value. They correspond to $5 \sigma$, where $\sigma$ is the standard deviation obtained via Monte-Carlo simulation of the experiment.}
    \label{table_certification}
\end{table}

To quantify high-dimensional entanglement, we measure the entanglement
of formation (EoF). The EoF is a measure of how many Bell states would need to be used in order to transform a single copy of our high-dimensional entangled state using only local operations and classical communication. As described in Ref.~\cite{schneeloch_quantifying_2018}, a lower bound of the EoF along the $X$-axis is expressed as follow
\begin{equation}
    E_x \geq -\text{log}_2 (e \Delta[x_1-x_2] \Delta[k_{x_1}+k_{x_2}]),
   \label{eq_eof_bound}
\end{equation}
where $\Delta[x_1-x_2]$ and $\Delta[k_{x_1}+k_{x_2}]$ are the uncertainty for measurements in the transformed variables $x_2-x_1$ and $k_{x_1}+k_{x_2}$, respectively. The same inequality exists for $E_y$, the lower bound of EoF along the $Y$-axis, after substituting $x_1-x_2 \rightarrow y_1-y_2$ and $k_{x_1}-k_{x_2} \rightarrow k_{y_1}-k_{y_2}$. Uncertainty values are estimated from the sum and minus-coordinate JPD projections shown in Figures~\ref{Figure2}.d.h, respectively, by Gaussian fitting. In our work, we find $-\text{log}_2 (e \Delta[x_1-x_2] \Delta[k_{x_1}+k_{x_2}] = 3.03(2)$ and $-\text{log}_2 (e \Delta[y_1-y_2] \Delta[k_{y_1}+k_{y_2}]) = 2.81(2)$. EoF values then provide lower bounds for entanglement dimensionality $d_x \geq 8$ ($X$-axis) and $d_y \geq 6$ ($Y$-axis) using the formula $d \geq 2 ^ {E}$~\cite{schneeloch_quantifying_2018}. The dimension of the state measured is thus higher than $14$.

We demonstrated the use of a single-photon sensitive time-stamping camera to quantify high-dimensional spatial entanglement without performing accidental background subtraction. Our approach improves on all previous camera-based techniques~\cite{tasca_imaging_2012,moreau_realization_2012,moreau_einstein-podolsky-rosen_2014,reichert_massively_2018,dabrowski_einsteinpodolskyrosen_2017,dabrowski_certification_2018,ndagano_imaging_2020,eckmann_characterization_2020}, in which such a post-processing step is required to demonstrate the presence of entanglement. Furthermore, by accurately detecting all output modes in parallel, we come closer to an ideal experimental configuration for certifying high-dimensional entanglement without assumptions, a situation that would be unattainable using single-outcome measurement approaches. From a practical point of view, we achieved high-dimensional entanglement quantification without accidental subtraction in $400$ seconds by detecting more than $6000$ spatial output modes  (i.e. illuminated pixels) in two bases, which is to our knowledge one of the fastest approach in term of acquisition time per number of modes. 

Nevertheless, our approach still needs improvement to reach assumptions-free high-dimensional entanglement certification. From the theory side, it is important to note that the entanglement criteria used in our study (i.e. equations~\eqref{epr_criterion} and~\eqref{eq_eof_bound}) do not enable \textit{certification} of high-dimensional entanglement in the strict sense of the term, because they use assumptions about the state e.g. purity and double-Gaussian approximation~\cite{schneeloch_introduction_2016,schneeloch_quantifying_2018}. In our work, we thus prefer to use the terms detection and quantification. To achieve proper certification, one would need to adapt more robust protocols, such as those developed for single-outcome measurements~\cite{bavaresco_measurements_2018}, to the case of camera detection. This is not straightforward, but could be done by specific design of the measurement bases~\cite{tasca_mutual_2018} and by using multi-plane light converters~\cite{labroille_efficient_2014}. From the technical side, we would like to point out that the image intensifier that we use has the quantum efficiency of about $20\%$ , which is insufficient to close the fair-sampling loophole~\cite{christensen_detection-loophole-free_2013}. The commercial image intensifiers do not achieve the quantum efficiency of more than $40\%$, so one would need to employ instead a single-photon sensitive sensor with higher quantum efficiency, for example, single photon avalanche detector, SPAD~\cite{lee_progress_2018}, or superconducting nanowire single-photon detectors, SNSPD,~\cite{wollman_kilopixel_2019}. These detectors would need to be combined with a data-driven readout as in \textit{Tpx3Cam}.

Furthermore, even if their rates are very low, we still detect accidental coincidences in the measured JPDs, i.e. approximately $1.4.10^{-4}$ accidental coincidences per pixel pair per second in the FF configuration, and $1.8.10^{-4}$ accidental coincidences per pixel pair per second in the NF configuration. These accidentals increase the measured uncertainties and therefore decreases the amount of entanglement dimension that can be quantified. They could be further reduced by diminishing the system losses e.g. using a type-II SPDC source to split photon paths with a polarising beam splitter, and by enhancing the effective camera temporal resolution~\cite{vidyapin_characterisation_2023}. 
\\\\
\textbf{Aknowledgements.} S.G. acknowledges funding from the European Research Council ERC Consolidator Grant (SMARTIES-724473). H.D. acknowledges funding from the European Research Council ERC Starting Grant (SQIMIC-101039375). P.S. acknowledges Centre of Advanced Applied Sciences (CZ.02.1.01/0.0/0.0/16-019/0000778), co-financed by the European Union, and A.N. acknowledges support by the BNL LDRD grant 22-22. We thank Rene Glazenborg and the company Photonis for loaning us the image intensifier.\\\\
\noindent \textbf{Authors contributions.} BC and H.D. performed the experiment. B.C., C.V. and H.D. analyzed the results. P.S and A.N. provided the $Tpx3cam$ and developed the centroiding and coincidence counting algorithms. H.D. and A.N. conceived the original idea. H.D. designed the experiment and supervised the project. All authors discussed the data and contributed to the manuscript.

\newpage

\clearpage
\section*{Supplementary document} 
\noindent \textbf{Details on the experimental apparatus.} 
The pump is a collimated continuous-wave laser at $405$ nm (Oxxius) with an output power of $50$ mW and a beam diameter of $0.8\pm 0.1$ mm. BBO crystal has dimensions $0.5 \times 5 \times 5$ mm and is cut for type I SPDC at $405$ nm with a half opening angle of $3$ degrees (Newlight Photonics). The crystal is slightly rotated around horizontal axis to ensure near-collinear phase matching of photons at the output (i.e. ring collapsed into a disk). A $650$ nm-cut-off long-pass filter is used to block pump photons after the crystals, together with a band-pass filter centered at $810 \pm 5$ nm. 

The $4f$ imaging system $f_1-f_2$ in Figure~\ref{Figure1}.a is represented by two lenses for clarity, but is in reality composed of a series of $8$ lenses with focal lengths $50$ mm - $150$ mm - $100$ mm - $200$ mm-$200$mm - $100$ mm - $75$ mm - $50$ mm. The first and the last lens are positioned at focal distances from the crystal and the plan P, and the distance between two lenses in a row equals the sum of their focal lengths. The other lenses have the following focal lengths: $f_3=30$mm, $f_4=150$mm. In the momentum basis (FF configuration), the system effective focal length is $75$ mm. In the position basis (NF configuration), the imaging system magnification is $10$. 

The camera is a \textit{Tpx3Cam} camera (Amsterdam Scientific Instruments) combined with an image intensifier (Cricket, from Photonis). The intensifier has a quantum efficiency of approximately $20\%$ at $810nm$. The camera has $256 \times 256$ pixels of $55\times55$ $\mu m$ each. Each pixel of the camera operates and is read out independently with time resolution of $1.56$ns and a dead time of about one microsecond. In practice, however, the timing dependence on the signal amplitude (timewalk effect) and small pixel to pixel time offsets reduce the effective temporal resolution to about $6$~ns~ (FWHM)~\cite{nomerotski_imaging_2019}, see also next section for more detail. 
\\\\
\noindent \textbf{Data processing.} Spatial joint probability distribution (JPD) measurement is achieved in three steps. (i) Data is acquired during $200$ seconds. The camera returns a list of hit pixels with spatial and temporal information. (ii) A centroiding algorithm processes the list to identify the true position for each photon. Indeed, each single photon at the input produces a cluster of hit pixels at the output due to the amplification process in the image intensifier and optical focusing convoluted with the physical pixel size. Our algorithm detects these clusters and reconstructs their centers as the amplitude weighted average over the whole cluster. Time of the pixel with the largest amplitude is used as measured time of the cluster. Timewalk correction is also obtained and applied at this stage. (iii) A pairing algorithm scans the list and selects only the events detected within a $6$~ns time window, i.e. the coincidence events. This value was chosen to optimize the signal to background ratio using the two photon time difference distribution. 
After this step, we know the number of coincidences measured per pairs of pixels, which therefore correspond to a sampling of all the JPD elements i.e. $\Gamma(x_1,y_1,x_2,y_2)$.

To visualize the measured JPDs, we use several types of projections, as shown in Figure~\ref{Figure2}. It includes marginal probability distributions $\Gamma_m({x_2},{y_2}) = \sum_{x_1,y_1} \Gamma({x_1},{y_2},x_1,y_1)$, conditional probability distributions $\Gamma({x_2},{y_2}|{x_1},{y_1}) = \Gamma({x_2},{y_2}|{x_1},{y_1}) / \Gamma_m({x_1},{y_1})$, minus-coordinate projections $\Gamma_-({x_1}-{x_2},{y_1}-{y_2}) = \sum_{x,y} \Gamma(x_1-x_2+x,y_1-y_2+y,x,y) $ and sum-coordinate projections $\Gamma_+({x_1}+{x_2},{y_1}+{y_2}) = \sum_{x,y} \Gamma(x_1+x_2-x,y_1+y_2-y,x,y)$. 
\\\\
\noindent \textbf{Uncertainty measurements and error analysis.} In our work, all measurable uncertainties, namely $\Delta[k_{x_2}|k_{x_1}]$ for all $k_{x_1}$,$\Delta[{x_2}|{x_1}]$ for all $x_1$, $\Delta[k_{y_2}|k_{y_1}]$ for all $k_{y_1}$,$\Delta[{y_2}|{y_1}]$ for all $y_1$, $\Delta[k_{x_2}+k_{x_1}]$,$\Delta[k_{y_2}+k_{y_1}]$,$\Delta[{x_2}-{x_1}]$ and $\Delta[{y_2}-{y_1}]$, were estimated from projections of the measured JPDs using bi-dimensional Gaussian fits i.e. $\exp(-x^2/2\Delta[x]^2-y^2/2\Delta[y]^2)$~\cite{schneeloch_introduction_2016}. Errors in the measured and calculated uncertainty values are written between parenthesis after each values and apply to the last digit of the value. They correspond to $5 \sigma$, where $\sigma$ is the standard deviation of the corresponding values estimated by performing Monte-Carlo simulations of the experiment. 

\bibliographystyle{abbrv}
\bibliography{Biblio2}

\end{document}